\documentclass[aps,prl,reprint,nofootinbib,superscriptaddress]{revtex4-2}
\usepackage{amsmath,amssymb,mathtools,bm,booktabs,mathrsfs,xcolor}
\usepackage{tikz}
\usetikzlibrary{arrows.meta,calc,decorations.pathmorphing,positioning}
\usepackage[colorlinks=true,linkcolor=blue!55!black,
 citecolor=blue!55!black,urlcolor=blue!55!black]{hyperref}
\definecolor{revisiongreen}{RGB}{0,85,45}
\definecolor{revisionred}{RGB}{180,0,0}
\newif\ifmarkchanges
\markchangestrue

\newcommand{\hplus}{+}
\newcommand{\hminus}{-}

\newcommand{\ET}{E_{\rm T}}
\newcommand{\bp}{\bm p}
\newcommand{\bk}{\bm k}
\newcommand{\bP}{\bm P}
\newcommand{\ang}[2]{\langle #1\,#2\rangle}
\newcommand{\df}[2]{D_{#1\mid #2}}
\newcommand{\Res}{\operatorname{Res}}

\begin{document}

\title{All-Plus QED Wavefunctions in de Sitter Space}
\author{Song He}
\email{songhe@itp.ac.cn}
\affiliation{New Cornerstone Laboratory, Institute of Theoretical Physics, Chinese Academy of Sciences, Beijing 100190, China}
\author{Jiajie Mei}
\email{j.mei@uva.nl}
\affiliation{Institute of Physics, University of Amsterdam, Amsterdam, 1098 XH, The Netherlands}
\author{Yuyu Mo}
\email{moyuyu@itp.ac.cn}
\affiliation{New Cornerstone Laboratory, Institute of Theoretical Physics, Chinese Academy of Sciences, Beijing 100190, China}

\begin{abstract}
We present a compact all-multiplicity formula for tree-level late-time wavefunction coefficients of scalar QED in four-dimensional de Sitter
space, with a conformally coupled charged scalar pair and positive-helicity
photons. The result admits both a recursive representation in terms
of shifted lower-point coefficients and an explicit finite sum involving
only kinematic factors.
The flat-space amplitudes vanish for two or more photons,
yet the de Sitter coefficients remain nonzero. These zeros enforce
cancellations of the total-energy residue and, through
factorization, of proper partial-energy residues involving
multi-photon subamplitudes.
Using a common-reference construction, we derive an exact lower-point
recursion from Abelian Ward identities. Conformal symmetry provides an
independent check: the partition sum solves the inhomogeneous Ward
identity, while the total-energy term is a homogeneous solution.
We also outline an analogous recursion in massless fermionic QED.
\end{abstract}

\maketitle

\section{Introduction}

A recurring lesson of the modern amplitudes program is that on-shell methods can organize entire families of amplitudes far more simply than their diagrammatic construction suggests~\cite{Parke:1986gb,Britto:2005fq,Dixon:2013amplitudes}. Late-time cosmological wavefunctions provide a natural setting in which to ask how far this simplicity extends beyond the flat-space $S$ matrix~\cite{Arkani-Hamed:2018kmz}. Their energy singularities encode bulk locality: at tree level, the leading total-energy singularity is controlled by the corresponding flat-space amplitude, while partial-energy singularities factorize into flat-space subamplitudes and complementary wavefunction data~\cite{Raju:2012flat,Benincasa:2018flat,Goodhew:2020optical,Baumann:2020dch}. Yet the full wavefunction contains information that need not be visible in these flat-space limits.

For certain classes of scalar wavefunctions, cosmological polytopes provide a geometric realization of this singularity and factorization structure~\cite{ArkaniHamed:2017polytope}. Comparable organizing principles for spinning correlators are less developed. Recent work on the cosmological Grassmannian shows that conformal symmetry and current conservation can be incorporated directly into a kinematic framework for massless spinning correlators, leading to striking simplifications at low multiplicity~\cite{Arundine:2026grassmannian}. This motivates the study of all-multiplicity fixed-helicity sectors, where helicity selection rules, factorization, and gauge symmetry can be brought together in a controlled setting.

The all-plus sector of scalar QED provides a particularly clean example. Tree amplitudes with a massless charged scalar pair and two or more positive-helicity photons vanish~\cite{Dixon:2013amplitudes}, whereas the corresponding late-time coefficients for a conformally coupled scalar pair on $\mathrm{dS}_4$ are nonzero. Their total-energy residue therefore vanishes, and factorization eliminates proper partial-energy residues whenever the associated flat-space subamplitude contains two or more photons. Hence only channels isolating a single cubic scalar--scalar--photon subdiagram can support proper partial-energy poles; residues in all additional channels present in individual diagrams must cancel in the full coefficient. The nonzero dS coefficient despite the vanishing flat-space amplitude demonstrates that the wavefunction contains genuinely new information beyond the flat-space S-matrix.

In this Letter, we derive a compact all-multiplicity formula for these
coefficients, expressing the $n$-photon wavefunction recursively in terms
of lower-point coefficients with shifted scalar momenta and an explicit
spinor-helicity term proportional to the total energy. The recursion
follows from a common-reference construction together with Abelian Ward
identities, while conformal Ward identities provide an independent check.
We also outline an analogous recursion in massless fermionic QED.

\section{All-multiplicity formula}

We work in the expanding Poincar\'e patch of $\mathrm{dS}_4$,
\begin{equation}
 ds^2=\frac{-d\eta^2+d\bm x^2}{H^2\eta^2},
 \qquad -\infty<\eta<0,
 \label{eq:ds-metric}
\end{equation}
and consider a conformally coupled complex scalar minimally coupled to an
Abelian gauge field. After the standard Weyl rescaling, the dynamics on
the Lorentzian conformal half-space are governed by
\begin{equation}
 S_\phi=\int_{\eta<0}d\eta\,d^3x\left[
 -\frac14F_{\mu\nu}F^{\mu\nu}
 -(\mathscr D_\mu\Phi)^*\mathscr D^\mu\Phi\right],
 \label{eq:scalar-action}
\end{equation}
with $\mathscr D_\mu=\partial_\mu-iA_\mu$ and mostly-plus signature. We
set the gauge coupling to $e=1$.

With Bunch--Davies initial conditions, fixed late-time boundary fields define
the wavefunctional
\begin{equation}
 \Psi_{\rm BD}[\varphi^*,\varphi,A_i]
 =\int_{\rm BD}^{\varphi^*,\varphi,A_i}
 \mathcal D\Phi\,\mathcal D\Phi^*\,\mathcal D A_\mu\,
 e^{iS_\phi}.
 \label{eq:BD-wavefunctional}
\end{equation}
Our bracket notation denotes the connected coefficients in the
boundary-field expansion of $-\log\Psi_{\rm BD}$ about vanishing fields.

We denote the charge-$-1$ and
charge-$+1$ scalar insertions by $O_u$ and $O_v$, respectively. At fixed
nonzero momentum, rotations about $\bk$ decompose a spin-one current into
helicity sectors $h=+1,0,-1$; we write
\begin{equation}
 J_i(\bk)=J^{-}(\bk)\,\xi_i^{+}(\bk)
 +J^{0}(\bk)\,\hat{k}_i
 +J^{+}(\bk)\,\xi_i^{-}(\bk).
 \label{eq:current-helicity-decomposition}
\end{equation}
The longitudinal component $J^0$ is fixed by the Ward identity. Choosing
$\bm\xi^\pm$ with $\bk\!\cdot\!\bm\xi^\pm=0$,
$\bm\xi^+\!\cdot\!\bm\xi^+=0$, and
$\bm\xi^+\!\cdot\!\bm\xi^-=1$, we project every photon onto
$h=+1$ and define
\begin{equation}
\Psi^\phi_{\mathcal P_n}(\bp_u,\bp_v)
:=
\left\langle
O_u(\bp_u)
\prod_{a=1}^{n}J_a^{+}(\bk_a)
O_v(\bp_v)
\right\rangle'.
\end{equation}
Here $\mathcal P_n=\{1,\ldots,n\}$,
$J_a^{\hplus}\equiv\xi_a^{\hplus\,i}J_i(\bk_a)$,
$k_a=|\bk_a|$, and $p_r=|\bp_r|$ for $r=u,v$. The prime removes the
spatial momentum-conserving delta function, and the external kinematics
obey
\begin{equation}
 \bp_{u}+\bp_{v}+\sum_{a=1}^{n}\bk_a=0,
 \qquad
 \ET={p_{u}+p_{v}}+\sum_{a=1}^{n}k_a .
 \label{eq:momentum-conservation}
\end{equation}
We normalize the scalar two-point coefficient as
$\Psi^\phi_\varnothing(\bp,-\bp)=p$.

We use spinors $\lambda_a,\bar\lambda_a$ \cite{Maldacena:2011nz} satisfying
\begin{equation}
 k^a_i\sigma^i_{\alpha\beta}
 =\lambda^a_{(\alpha}\bar\lambda^a_{\beta)},
 \qquad
 \ang{a}{\bar a}=-2k_a,
 \qquad k_a=|\bk_a|.
 \label{eq:momentum-spinors}
\end{equation}
We use the same convention for scalar momenta. Angle brackets denote
antisymmetric spinor contractions. For either endpoint $r\in\{u,v\}$, define
\begin{equation}
\begin{aligned}
 \df{a}{r}&:=\frac{\ang{\bar a}{r}}
 {2k_a\ang{a}{r}}, \quad
 \bk_S:=\sum_{a\in S}\bk_a,
 \quad \bP_{r S}:=\bp_{r}+\bk_S.
\end{aligned}
 \label{eq:endpoint-factors}
\end{equation}
Here $P_{rS}=|\bP_{rS}|$ is the energy of a merged endpoint.

Our central result is the following all-multiplicity relation. For
$n\geq1$,
\begin{multline}
 \Psi^\phi_{\mathcal P_n}(\bp_{u},\bp_{v})
 =\frac{\ET\,\ang{u}{v}^{n}}
 {2\prod_{a=1}^{n}\ang{u}{a}\ang{v}{a}}
 \\[-1mm]
 {}+\smashoperator[r]{\sum_{\substack{L\sqcup C\sqcup R=\mathcal P_n\\
                   L\cup R\ne\varnothing}}}
 (-1)^{|R|+1}
 \prod_{a\in L}\df{a}{u}\,
 \prod_{b\in R}\df{b}{v}\,
 \Psi^\phi_C(\bP_{u L},\bP_{v R}).
 \label{eq:scalar-master}
\end{multline}
The partition sum has a direct recursive interpretation. Each partition
$L\sqcup C\sqcup R=\mathcal P_n$ assigns photons either to the left
endpoint, to the lower-point coefficient, or to the right endpoint. The
explicit $D$ factors are evaluated using the original endpoint momenta,
whereas $\Psi_C^\phi$ is evaluated on the corresponding merged momenta.
Since $L\cup R\ne\varnothing$, each coefficient in the sum has fewer
photons, and iteration terminates at $\Psi^\phi_\varnothing$.

The recursion can be iterated until no lower-point coefficient remains.
Here $|P_{rS}\rangle$ denotes the spinor of $\bP_{rS}$, and
$D_{a\mid P_{rS}}$ is the endpoint factor~\eqref{eq:endpoint-factors}
evaluated at this momentum. We write
$D_{B\mid x}:=\prod_{a\in B}D_{a\mid x}$ for either an original or
merged endpoint $x$. Define
\begin{multline}
 \mathcal D_{L\mid r}:=\sum_{(B_1,\ldots,B_p)}(-1)^p
 D_{B_1\mid r}\,D_{B_2\mid P_{rB_1}}
 \\[-1mm]
 {}\times\cdots\times D_{B_p\mid P_{rB_1\cdots B_{p-1}}},
 \label{eq:scalar-absorption-factor}
\end{multline}
where the sum includes all numbers and orders of nonempty groups with
$B_1\sqcup\cdots\sqcup B_p=L$, and
$\mathcal D_{\varnothing\mid r}=1$.
Each group's factors use the endpoint shifted by the earlier groups;
multiple set subscripts denote their union
[Fig.~\ref{fig:all-multiplicity-mechanism}(b)]. Then
\begin{multline}
 \Psi^\phi_{\mathcal P_n}(\bp_u,\bp_v)
 =\sum_{L\sqcup C\sqcup R=\mathcal P_n}
 (-1)^{|R|}\mathcal D_{L\mid u}\mathcal D_{R\mid v}
 \\[-1mm]
 {}\times\frac{\bigl(P_{uL}+\sum_{c\in C}k_c+P_{vR}\bigr)
 \ang{P_{uL}}{P_{vR}}^{|C|}}
 {2\prod_{c\in C}\ang{P_{uL}}{c}\ang{P_{vR}}{c}}.
 \label{eq:scalar-unrolled}
\end{multline}
For $C=\varnothing$, the last factor reduces to $P_{uL}=P_{vR}$.

Additional checks include comparison with explicit Feynman-rule
calculations and soft-photon limits~\cite{Mei:2025soft} up to 10 points numerically. We next examine the conformal Ward
identity and the pole structure.

\begin{figure}[htbp]
\centering
\begingroup
\colorlet{unrolledcolor}{black}
\color{unrolledcolor}
\hypersetup{linkcolor=unrolledcolor}
\begin{tikzpicture}[
  x=1cm,y=1cm,font=\small,color=unrolledcolor,
  every node/.style={text=unrolledcolor},
  charged/.style={line width=1.15pt},
  endpoint/.style={circle,draw,fill=white,inner sep=1.8pt,line width=0.9pt},
  cluster/.style={circle,draw,fill=unrolledcolor!8,minimum size=0.62cm,
    inner sep=0pt,line width=0.9pt},
  core/.style={draw,rounded corners=1.5pt,fill=unrolledcolor!8,
    minimum width=1.25cm,minimum height=0.72cm,line width=0.9pt},
  photon/.style={decorate,decoration={snake,amplitude=0.45mm,
    segment length=2mm,pre length=0.5mm,post length=0.5mm},
    line width=0.75pt}
]
\node[font=\small\bfseries] at (-0.35,1.65) {(a)};
\draw[charged] (0.10,0)--(7.30,0);
\node[cluster] (cl) at (1.05,0) {};
\node[cluster] (cr) at (6.35,0) {};
\node[core] (cc) at (3.70,0) {};
\draw[photon] (0.30,1.25)--(0.85,0.30);
\draw[photon] (1.80,1.25)--(1.25,0.30);
\node at (1.05,1.10) {$\cdots$};
\node at (1.05,1.65) {$L$};
\draw[photon] (5.60,1.25)--(6.15,0.30);
\draw[photon] (7.10,1.25)--(6.55,0.30);
\node at (6.35,1.10) {$\cdots$};
\node at (6.35,1.65) {$R$};
\draw[photon] (3.20,1.25)--(3.40,0.37);
\draw[photon] (4.20,1.25)--(4.00,0.37);
\node at (3.70,1.10) {$\cdots$};
\node at (3.70,1.65) {$C$};
\node[below=3pt] at (0.10,0) {$u$};
\node[below=3pt] at (7.30,0) {$v$};
\node[below=3pt] at (2.30,0) {$\bP_{uL}$};
\node[below=3pt] at (5.10,0) {$\bP_{vR}$};
\node[font=\scriptsize] at (1.05,-0.85) {$\mathcal D_{L\mid u}$};
\node[font=\scriptsize] at (6.35,-0.85) {$(-1)^{|R|}\mathcal D_{R\mid v}$};
\node[font=\scriptsize] at (3.70,-1.45)
  {$\dfrac{(P_{uL}+\sum_{c\in C}k_c+P_{vR})\ang{P_{uL}}{P_{vR}}^{|C|}}{2\prod_{c\in C}\ang{P_{uL}}{c}\ang{P_{vR}}{c}}$};
\draw[line width=0.4pt,densely dotted] (3.70,-0.42)--(3.70,-1.02);
\begin{scope}[yshift=-3.65cm]
\node[font=\small\bfseries] at (-0.35,1.65) {(b)};
\draw[charged] (0.10,0)--(7.30,0);
\node[endpoint] (b1) at (1.30,0) {};
\node[endpoint] (b2) at (3.10,0) {};
\node[endpoint] (bp) at (5.70,0) {};
\node at (4.40,0.30) {$\cdots$};
\foreach \x/\lab in {1.30/B_1,3.10/B_2,5.70/B_p}{
  \draw[photon] (\x-0.55,1.20)--(\x,0);
  \draw[photon] (\x+0.55,1.20)--(\x,0);
  \node at (\x,1.05) {$\cdots$};
  \node at (\x,1.55) {$\lab$};
}
\node[below=3pt] at (0.10,0) {$u$};
\node[below=3pt] at (2.20,0) {$\bP_{uB_1}$};
\node[below=3pt] at (7.05,0) {$\bP_{uL}$};
\node[font=\scriptsize] at (1.30,-0.85) {$\df{B_1}{u}$};
\node[font=\scriptsize] at (3.10,-0.85) {$\df{B_2}{P_{uB_1}}$};
\node[font=\scriptsize] at (5.70,-0.85) {$\df{B_p}{P_{uB_1\cdots B_{p-1}}}$};
\end{scope}
\end{tikzpicture}
\caption{(a) The nonrecursive partition sum in
Eq.~\eqref{eq:scalar-unrolled}. The endpoint factors absorb $L$ and $R$;
the photons in $C$ enter the displayed kinematic factor.
(b) One term of $\mathcal D_{L\mid u}$: the endpoint absorbs the ordered groups
$B_1,\ldots,B_p$, with each $D$ factor evaluated before its group is
absorbed. Sum over all such groupings and orders with sign $(-1)^p$.
Solid lines track endpoint momentum shifts; wavy lines denote photons.}
\label{fig:all-multiplicity-mechanism}
\endgroup
\end{figure}

\section{Symmetry and pole structure}
\subsection{Conformal Ward identity}

De Sitter symmetry provides an independent consistency check of
Eq.~\eqref{eq:scalar-master} and distinguishes the roles of its two terms.
In spinor variables~\cite{Baumann:2020dch}, with
$\mathcal K^i=2(\sigma^i)_\alpha{}^\beta
\sum_j\partial_{\lambda_{j\alpha}}\partial_{\bar\lambda_j{}^\beta}$,
the wavefunction coefficient obeys the inhomogeneous Ward identity
\begin{align}
 \mathcal K^i\Psi^\phi_{\mathcal P_n}
 ={}&2\sum_{a\in\mathcal P_n}\frac{\xi_a^{\hplus,i}}{k_a^2}
 \Big[
 \Psi^\phi_{\mathcal P_n\setminus\{a\}}
 (\bp_u,\bP_{v\{a\}})
 \nonumber\\[-1mm]
 &\hspace{5.2em}-
 \Psi^\phi_{\mathcal P_n\setminus\{a\}}
 (\bP_{u\{a\}},\bp_v)
 \Big].
 \label{eq:scalar-CWI}
\end{align}
The explicit term proportional to $\ET$ is annihilated by
$\mathcal K^i$,
\begin{equation}
 \mathcal K^i\!\left[
 \frac{\ET\,\ang{u}{v}^{n}}
 {2\prod_{a=1}^{n}\ang{u}{a}\ang{v}{a}}
 \right]=0.
\end{equation}
The partition sum supplies the inhomogeneous source. Acting
on its endpoint factors and using the lower-point conformal Ward
identities reproduces the right-hand side of Eq.~\eqref{eq:scalar-CWI};
the required intertwining identity is given in
Appendix~\ref{app:symmetry}.

\subsection{Pole structure and low-point examples}

\paragraph{Pole constraints.}
The flat-space zero constrains the residues of the wavefunction's energy
singularities.
For a massless charged scalar pair,
\begin{equation}
 \mathcal A^{\rm tree}_{m+2}
 (\phi,\phi^*,a_1^+,\ldots,a_m^+)=0,
 \qquad m\geq2,
 \label{eq:scalar-amplitude-selection}
\end{equation}
in scalar QED~\cite{Dixon:2013amplitudes}. 
Using the partition notation of Eq.~\eqref{eq:scalar-master}, take
$L,C,R\neq\varnothing$ and isolate the central block $C$ by cutting
the two adjacent scalar lines. The associated partial energy is
\[
E_{P_{uL}\mid C\mid P_{vR}}
:=P_{uL}+\sum_{c\in C}k_c+P_{vR}.
\]
At a generic point on this analytically continued factorization locus,
away from other singularities, the residue is
~\cite{Benincasa:2018flat,Baumann:2020dch}

\begin{equation}
 \underset{E_{P_{uL}\mid C\mid P_{vR}}=0}{\Res}\,
 \Psi^\phi_{\mathcal P_n}
 =\frac{\operatorname{Disc}_{P_{uL}^{2}}\Psi^\phi_L}
 {2P_{uL}}\,
 \mathcal A^{\rm flat}_{C}\,
 \frac{\operatorname{Disc}_{P_{vR}^{2}}\Psi^\phi_R}
 {2P_{vR}}.
 \label{eq:all-plus-factorization}
\end{equation}
Here $\mathcal A^{\rm flat}_{C}
=\mathcal A^{\rm tree}_{|C|+2}
(\phi,\phi^*,\{a_c^{\hplus}\}_{c\in C})$ has scalar momenta
$\bP_{uL}$ and $\bP_{vR}$. The complementary coefficients
$\Psi_L^\phi$ and $\Psi_R^\phi$ have scalar pairs
$(\bp_u,-\bP_{uL})$ and $(-\bP_{vR},\bp_v)$, respectively.
We use $\operatorname{Disc}_{P^2}F:=F(P)-F(-P)$, which flips the
cut scalar energy while keeping its spatial momentum fixed; the
$1/(2P)$ factors are the scalar cut-line normalizations
\cite{He:2026cuts}. For endpoint subdiagrams, the corresponding
factorization involves a single cut.

Equation~\eqref{eq:scalar-amplitude-selection} therefore removes the
residue whenever the flat-space subamplitude contains two or more photons.
The only proper partial-energy poles not excluded by this rule are those
of a single cubic scalar--scalar--photon subdiagram. Applying the same
argument to the full process also removes the total-energy residue for
$n\geq2$. These statements constrain the complete coefficient: the
forbidden diagrammatic poles must cancel in the sum organized by
Eq.~\eqref{eq:scalar-master}.

\paragraph{Low-point examples.}
The first cases show how these constraints emerge from the recursive
formula. With one photon, Eq.~\eqref{eq:scalar-master} gives
\begin{equation}
\begin{aligned}
 \Psi^\phi_{\{1\}}
 &=\frac{\ET\ang{u}{v}}{2\ang{u}{1}\ang{v}{1}}
 -\df{1}{u}p_v+\df{1}{v}p_u.
\end{aligned}
 \label{eq:scalar-three-point}
\end{equation}

Four points are the first case in which the flat-space zero constrains
the total-energy singularity. Substituting the three-point result into the partition sum and using momentum conservation, and Schouten's identity gives
(see Appendix~\ref{app:four-point-specialization})
\begin{equation}
\begin{aligned}
\Psi_{\{1,2\}}^{\phi}
={}&D_{1|u}D_{2|u}(\tfrac{\ET}{2}-p_v)
+D_{1|v}D_{2|v}(\tfrac{\ET}{2}-p_u)\\
&+D_{1|u}D_{2|v}(\tfrac{\ET}{2}-P_{u1})
+D_{2|u}D_{1|v}(\tfrac{\ET}{2}-P_{u2}).
\end{aligned}
 \label{eq:scalar-four-point}
\end{equation}
The coefficient has no $1/\ET$ pole: the total-energy residues of the
two scalar-exchange orderings and the quartic interaction cancel, as illustrated in
Fig.~\ref{fig:four-point-cancellation}.

\begin{figure}[t]
\centering
\begin{tikzpicture}[
  x=1cm,
  y=1cm,
  font=\scriptsize,
  charged/.style={line width=0.9pt},
  vertex/.style={circle,fill=black,inner sep=1.25pt},
  endpoint/.style={circle,draw,fill=white,inner sep=1.15pt,line width=0.7pt},
  photon/.style={decorate,decoration={snake,amplitude=0.42mm,
    segment length=1.8mm,pre length=0.4mm,post length=0.4mm},line width=0.7pt}
]

\draw[charged] (0.10,0)--(1.90,0);
\node[endpoint] at (0.10,0) {};
\node[endpoint] at (1.90,0) {};
\node[vertex] at (0.70,0) {};
\node[vertex] at (1.30,0) {};
\draw[photon] (0.70,0.03)--(0.70,0.92);
\draw[photon] (1.30,0.03)--(1.30,0.92);
\node at (0.70,1.08) {$a^{\hplus}$};
\node at (1.30,1.08) {$b^{\hplus}$};
\node at (1.00,-0.30) {$(a,b)$};

\node[font=\normalsize] at (2.22,0.34) {$+$};

\draw[charged] (2.54,0)--(4.34,0);
\node[endpoint] at (2.54,0) {};
\node[endpoint] at (4.34,0) {};
\node[vertex] at (3.14,0) {};
\node[vertex] at (3.74,0) {};
\draw[photon] (3.14,0.03)--(3.14,0.92);
\draw[photon] (3.74,0.03)--(3.74,0.92);
\node at (3.14,1.08) {$b^{\hplus}$};
\node at (3.74,1.08) {$a^{\hplus}$};
\node at (3.44,-0.30) {$(b,a)$};

\node[font=\normalsize] at (4.66,0.34) {$+$};

\draw[charged] (4.98,0)--(6.78,0);
\node[endpoint] at (4.98,0) {};
\node[endpoint] at (6.78,0) {};
\node[vertex,inner sep=1.75pt] at (5.88,0) {};
\draw[photon] (5.85,0.03)--(5.48,0.92);
\draw[photon] (5.91,0.03)--(6.28,0.92);
\node at (5.43,1.08) {$a^{\hplus}$};
\node at (6.33,1.08) {$b^{\hplus}$};
\node at (5.88,-0.30) {quartic};

\node[font=\small] at (3.44,-0.72)
  {$\displaystyle \Res_{\ET=0}\Psi^\phi_{\{a,b\}}=0$};
\end{tikzpicture}
\caption{The four-point scalar coefficient in physical gauge. The two
scalar-exchange orderings and the quartic contact graph combine into
Eq.~\eqref{eq:scalar-four-point}; their total-energy residues cancel
in the sum. Internal energy denominators are suppressed in the drawing.}
\label{fig:four-point-cancellation}
\end{figure}

At five points the recursion contains four-point coefficients with one
shifted charged endpoint. Six points first allow a
four-point coefficient with both endpoints shifted. For example,
$L=\{1\}$, $C=\{2,3\}$, $R=\{4\}$ contributes
$\df{1}{u}\df{4}{v}\,\Psi^\phi_{\{2,3\}}(\bP_{u1},\bP_{v4})$.
The internal $D$ factors of this four-point coefficient use merged
endpoint spinors; the explicit prefactors $D_{1|u}D_{4|v}$ use the
original endpoint momenta.

The same six-point kinematics illustrate the selection of factorization
channels. For the photon sequence $(1,2,3,4)$ along a cubic scalar line
from $u$ to $v$, the one-photon subdiagrams have energies
$Y_2=P_{u1}+k_2+P_{u12}$ and $Y_3=P_{u12}+k_3+P_{u123}$.
Their combined two-photon subdiagram has energy
$X=P_{u1}+k_2+k_3+P_{u123}$.
Here $P_{u12}\equiv P_{u\{1,2\}}$, with analogous shorthand for
other multiple-label subscripts. The residue at $X=0$ vanishes in the
full coefficient, whereas the one-photon channels $Y_2$ and $Y_3$ are
not excluded. The full Abelian coefficient includes all photon orderings.

The bracket denominators in Eq.~\eqref{eq:scalar-master} encode these
energy loci in spinor variables. For example, the $Y_2$ channel obeys
\begin{equation}
\begin{aligned}
 (P_{u1}+k_2+P_{u12})(P_{u1}+k_2-P_{u12})=\langle P_{u1}\,2\rangle
 \langle \bar P_{u1}\,\bar 2\rangle .
\end{aligned}
 \label{eq:spinor-energy-pole}
\end{equation}
The bracket labels $P_{u1}$ and $\bar P_{u1}$ denote the spinors of
$\bP_{u1}$. On complexified kinematics, $Y_2$ times its folded partner
$P_{u1}+k_2-P_{u12}$ thus splits into a product of spinor brackets.
The three-point seed
$\Psi^\phi_{\{a\}}=D_{a|u}(k_a+p_u-p_v)$ illustrates the role of the
numerator: rationalizing its bracket denominator cancels the folded
factor $k_a+p_u-p_v$, leaving the physical total-energy denominator
$k_a+p_u+p_v$.


\section{Ward-identity derivation}
To derive the recursion, we replace the physical positive-helicity polarizations
by common-reference auxiliary vectors and apply the Abelian Ward identity
to their longitudinal difference.

With the spinor convention in Eq.~\eqref{eq:momentum-spinors}, the
transverse helicity states can be written as
\begin{equation}
 \xi^{\hplus}_{a,\alpha\beta}
 =
 \frac{\bar\lambda_{a,\alpha}\bar\lambda_{a,\beta}}{2k_a},
 \qquad
 \xi^{\hminus}_{a,\alpha\beta}
 =
 \frac{\lambda_{a,\alpha}\lambda_{a,\beta}}{2k_a}.
 \label{eq:helicity-polarizations}
\end{equation}

Using $u$ as a common reference, define the auxiliary vector%
\footnote{This choice is motivated by the common-reference gauge used
for all-plus amplitudes~\cite{Dixon:2013amplitudes}.}
\begin{equation}
 \epsilon_{a,\alpha\beta}(u)
 :=\frac{\lambda_{u,(\alpha}\bar\lambda_{a,\beta)}}{\ang{u}{a}}
 =\xi^{\hplus}_{a,\alpha\beta}-\df{a}{u}k_{a,\alpha\beta}.
 \label{eq:gauge-shift}
\end{equation}
The second equality follows from Schouten's identity and isolates the
longitudinal term to which the Abelian Ward identity applies.

Let $\Psi^{\rm CR}_{\mathcal P_n}(u)$ denote the coefficient with every
photon contracted with $\epsilon_a(u)$, keeping
$J_{a,\alpha\beta}$ open before contraction. Expanding
$\epsilon_a(u)=\xi_a^+-D_{a|u}k_a$ for every photon gives the exact
subset identity
\begin{equation}
\begin{aligned}
 \Psi^\phi_{\mathcal P_n}
 ={}&\Psi^{\rm CR}_{\mathcal P_n}(u)\\
 &+\sum_{\varnothing\ne S\subseteq\mathcal P_n}
 (-1)^{|S|+1}
 \prod_{a\in S}\left(\df{a}{u}k_a^{\alpha_a\beta_a}\right)\\
 &\qquad\times
 \left\langle O_u\prod_{a\in S}J_{a,\alpha_a\beta_a}
 \prod_{b\in\mathcal P_n\setminus S}J_b^+O_v\right\rangle'.
\end{aligned}
 \label{eq:longitudinal-subset-expansion}
\end{equation}

The exceptional three-point auxiliary contraction supplies the seed of
the recursion. The cubic interaction gives (see Appendix~\ref{app:vertices})
\begin{equation}
\left\langle O_uJ_{a,\alpha\beta}O_v\right\rangle'
=\frac{1}{2\ET}\left[
 p_{u,\alpha\beta}-p_{v,\alpha\beta}
 -\frac{p_u-p_v}{k_a}k_{a,\alpha\beta}\right].
\label{eq:three-point-open-current}
\end{equation}
Contracting with the auxiliary vector gives
$\Psi^{\rm CR}_{\{a\}}(u)=k_a\df{a}{u}$. Combining this with the
three-point Abelian Ward identity,
$
 k_a^{\alpha\beta}\left\langle O_uJ_{a,\alpha\beta}O_v\right\rangle'=p_u-p_v,
$
we obtain
\begin{equation}
\begin{aligned}
 \Psi^\phi_{\{a\}}
 &=\df{a}{u}(k_a+p_u-p_v)\\
 &=
 \frac{\ET\ang{u}{v}}{2\ang{u}{a}\ang{v}{a}}
 -\df{a}{u}p_v+\df{a}{v}p_u.
\end{aligned}
 \label{eq:three-point-gauge-change}
\end{equation}
The endpoint-symmetric form follows by antisymmetrizing under exchange of the oppositely
charged endpoints,
$\Psi^\phi_{\{a\}}=\tfrac12[\Psi^\phi_{\{a\}}-(u\leftrightarrow v)]$,
and using Schouten's identity,
$
 \df{a}{u}-\df{a}{v}
 =\frac{\ang{u}{v}}{\ang{u}{a}\ang{v}{a}}.
$
This reproduces the one-photon case of Eq.~\eqref{eq:scalar-master}.

From four points onward the all-auxiliary contraction vanishes,
$\Psi^{\rm CR}_{\mathcal P_n}(u)=0$, as shown diagrammatically in Appendix~\ref{app:higher-point}. Each tree diagram begins at \(u\) with either a cubic or a quartic interaction, and contraction with the auxiliary vectors annihilates the corresponding vertex.
The physical coefficient is therefore reconstructed
entirely from terms containing at least one longitudinal photon momentum in
Eq.~\eqref{eq:longitudinal-subset-expansion}. For a
single contracted photon $a$, the Abelian Ward identity reads
\begin{equation}
\begin{aligned}
 k_a^{\alpha\beta}
 \langle O_uJ_{a,\alpha\beta}\cdots O_v\rangle'
 ={}&\Psi^\phi_{\mathcal P_n\setminus\{a\}}
 (\bp_u,\bP_{v\{a\}})\\
 &-\Psi^\phi_{\mathcal P_n\setminus\{a\}}
 (\bP_{u\{a\}},\bp_v).
\end{aligned}
\label{eq:abelian-ward-main}
\end{equation}
The ellipsis denotes the remaining physical-helicity currents. Each
contraction removes one photon and transfers its momentum to a charged
endpoint, with the relative sign fixed by the opposite charges. Iterating
Eq.~\eqref{eq:abelian-ward-main} gives an exact lower-point recursion
that, together with the three-point seed, terminates at the scalar
two-point coefficient.

Appendix~\ref{app:higher-point} presents the partition identity relating
this recursion to the symmetric expression in
Eq.~\eqref{eq:scalar-master}, with the reference endpoint kept fixed.
The zero- and one-photon auxiliary contributions account for the explicit
total-energy term.

\section{Fermionic QED}

We outline an analogous construction for massless fermionic QED. For
external helicities
$\bar\chi_u^+$ and $\chi_v^-$ and positive-helicity photons, define
\begingroup
\small
\begin{equation}
 \Psi^{\chi,+-}_{\mathcal P_n}(\bp_u,\bp_v)
 :=\left\langle
 \bar\chi_u^+(\bp_u)\prod_{a=1}^nJ_a^+(\bk_a)
 \chi_v^-(\bp_v)\right\rangle'.
 \label{eq:fermion-definition}
\end{equation}
\endgroup
For this helicity choice, the common-reference contraction with
$\epsilon_a(v)$ vanishes already for $n\geq1$, including at three
points. The same Abelian Ward reduction directly gives the recursion
\begingroup
\small
\begin{equation}
\begin{aligned}
 \Psi^{\chi,+-}_{\mathcal P_n}
 ={}&\sum_{\substack{L\sqcup C\sqcup R=\mathcal P_n\\L\cup R\ne\varnothing}}
 (-1)^{|R|+1}\prod_{a\in L}\df{a}{v}\prod_{b\in R}\df{b}{v}\\[-1mm]
 &\qquad\times\Psi^{\chi,+-}_C(\bP_{uL},\bP_{vR}).
\end{aligned}
 \label{eq:fermion-master}
\end{equation}
\endgroup
The sets $L,C,R$ and merged momenta have the same meaning as in
Eq.~\eqref{eq:scalar-master}. Both endpoint assignments carry $D_{a|v}$
because the common-reference vector is defined using the original endpoint
$v$. As in the scalar recursion, the momentum shifts act inside the
lower-point object; here that object retains open fermion indices, while
the external helicity spinors remain tied to the original momenta:
\begingroup
\small
\begin{equation}
 \Psi^{\chi,+-}_C(\bP_{uL},\bP_{vR})
 :=\bar\zeta_u^+\bm\Psi^\chi_C(\bP_{uL},\bP_{vR})\zeta_v^-.
 \label{eq:fermion-merged-coefficient}
\end{equation}
\endgroup
Iteration terminates at the corresponding two-point response. In this
helicity sector, the fermionic recursion has the same Abelian
endpoint-shift structure, with no separate term proportional to $\ET$. We have verified that our formula reproduces the four-point results of Ref.~\cite{Chen:2025fermionicflat} and agrees numerically with explicit Feynman diagram computations up to ten points.
\section{Discussion and outlook}

Organizing the coefficient through shifts of two charged endpoints
exposes a simplicity obscured by individual diagrams, recalling the role
of the Parke--Taylor formula~\cite{Parke:1986gb}. Here a compact
all-multiplicity structure persists even when the flat-space amplitude
vanishes. 
The all-plus sector thus illustrates how a nonzero cosmological wavefunction
retains information beyond its vanishing flat-space limit.

Factorization restricts the allowed residues, while the Ward
recursion with its seed reconstructs the coefficient. The explicit
total-energy term is homogeneous under the conformal Ward identity and
is not determined by that identity alone.

Whether a comparable organization exists beyond the all-plus Abelian
sector remains open. Mixed photon helicities need not preserve the
vanishing common-reference contraction, while non-Abelian theories
introduce gauge-boson self-interactions. Explicit higher-point Yang--Mills
results~\cite{Albayrak:2018higher,He:2026cuts,Gomez:2026gluon} and
four-point all-plus expressions~\cite{Armstrong:2020ck,Huang:2026supergrassmannian}
provide starting points for exploring an analogous
all-multiplicity organization.

A separate question is whether the restricted pole pattern admits a
geometric description appropriate to spinning
wavefunctions
~\cite{ArkaniHamed:2017polytope,Arundine:2026grassmannian}. The all-plus sector provides a simple all-multiplicity setting in which
to explore how helicity-induced zeros and gauge symmetry are encoded
geometrically.

\section*{Acknowledgement}
\begin{acknowledgments}
We thank Chandramouli Chowdhury and Yu-tin Huang for useful discussions
and comments.
S.H. and Y.M. are supported by the National Natural Science Foundation
of China under Grant Nos.~12225510 and 12447101, and by the New
Cornerstone Science Foundation.
J.M. is supported by the European Union (ERC, UNIVERSE PLUS, 101118787).
Views and opinions expressed are, however, those of the authors only and
do not necessarily reflect those of the European Union or the European
Research Council Executive Agency. Neither the European Union nor the
granting authority can be held responsible for them. OpenAI's GPT-5.6 Sol and GPT-6 models assisted with derivations, computational checks, code development, and manuscript preparation; all AI-assisted material and scientific conclusions were reviewed and independently verified by the authors, who take full responsibility for the content.
\end{acknowledgments}

\appendix
\setcounter{secnumdepth}{2}

\onecolumngrid

\section{External photon waves and interaction vertices}
\label{app:vertices}

Starting from the scalar-QED action on the Lorentzian conformal half-space,
$S_\phi=\int_{\eta<0}d\eta,d^3x[-\tfrac14F_{\mu\nu}F^{\mu\nu}-(\mathscr D_\mu\Phi)^*\mathscr D^\mu\Phi]$,
with $\mathscr D_\mu=\partial_\mu-iA_\mu$, one obtains
$\partial_\mu F^{\mu\nu}=j_{\rm bulk}^{\nu}$ with
$j_{\rm bulk}^{\nu}=i[\Phi^*\mathscr D^\nu\Phi-(\mathscr D^\nu\Phi)^*\Phi]$.
Using boundary translation invariance, we consider a Fourier mode
$e^{i\bk\cdot\bm x}$,
\begin{equation}
\begin{aligned}
(\partial_\eta^2+k^2)A_i&=ik_i\partial_\eta A_\eta+k_i(\bk\cdot\bm A)-j_{{\rm bulk},i},\\[-1mm]
k^2A_\eta+i\partial_\eta(\bk\cdot\bm A)&=j_{\rm bulk}^{\eta}.
\end{aligned}
\label{eq:appA-Maxwell}
\end{equation}
The second line is the Gauss constraint. In the presence of a bulk source,
the relation between $A_\eta$ and the longitudinal spatial component is
modified and is not an identity for a generic interacting or off-shell gauge
field. When considering an external photon leg, the gauge field obeys the
free bulk equation with $j_{\rm bulk}^{\mu}=0$. Therefore,
\begin{equation}
A_{a,\eta}=\frac{k_{a,i}A_{a,i}}{k_a}.
\label{eq:external-transversality}
\end{equation}
We can now determine the bulk-to-boundary propagator by imposing the Lorenz
gauge $\partial_\mu A^\mu=0$. Using Eq.~\eqref{eq:external-transversality},
\begin{equation}
\begin{aligned}
A_j(\eta,k) = e^{ik\eta}A_j^{(0)}(k), \qquad A_\eta(\eta,k) = e^{ik\eta}\hat{k}^i A_i^{(0)}(k).
\end{aligned}
\end{equation}
By expanding
$-(\mathscr D_\mu\Phi)^*\mathscr D^\mu\Phi=-\partial_\mu\Phi^*\partial^\mu\Phi
+iA^\mu(\partial_\mu\Phi^*)\Phi-iA^\mu\Phi^*\partial_\mu\Phi
-A_\mu A^\mu\Phi^*\Phi$
and using Eq.~\eqref{eq:external-transversality}, we identify the following
effective vertices:
\begin{equation}
\begin{aligned}
V_{3,\alpha\beta}
&=(p'-p)_{\alpha\beta}
-i\frac{k_{a,\alpha\beta}}{k_a}
(\overrightarrow{\partial}_\eta-\overleftarrow{\partial}_\eta),\\[-1mm]
(V_4)_{\alpha\beta,\gamma\delta}(a,b)
&=i\left(\mathbb I_{\alpha\beta,\gamma\delta}
+\frac{k_{a,\alpha\beta}k_{b,\gamma\delta}}{2k_a k_b}\right).
\end{aligned}
\label{eq:effective-cubic-rule}
\end{equation}
These effective vertex rules apply only when the gauge fields entering them
are external legs directly attached to the boundary. Here $\bp'+\bp+\bk_a=0$
and $F(\overrightarrow{\partial}_\eta-\overleftarrow{\partial}_\eta)H=FH'-F'H$.
The quartic spatial tensor is initially proportional to
$\delta_{ij}-k_{a,i}k_{b,j}/(k_a k_b)$.
For the $\delta_{ij}$ term,
$\frac14(-2i)\sigma_{\alpha\beta}^{i}\delta_{ij}\sigma_{\gamma\delta}^{j}
=-\frac{i}{2}\sigma_{\alpha\beta}^{i}\sigma_{\gamma\delta}^{i}
=i,\mathbb I_{\alpha\beta,\gamma\delta}$,
where we used the Pauli-matrix completeness relation
$\sigma_{\alpha\beta}^{i}\sigma_{\gamma\delta}^{i}
=-(\epsilon_{\alpha\gamma}\epsilon_{\beta\delta}
+\epsilon_{\alpha\delta}\epsilon_{\beta\gamma})$.
Likewise, the $k_{a,i}k_{b,j}/(k_a k_b)$ term gives
$i,k^a_{\alpha\beta}k^b_{\gamma\delta}/(2k_a k_b)$,
yielding the second line of Eq.~\eqref{eq:effective-cubic-rule}.
The resulting three-point contraction is the auxiliary block used in the
recursion, Eq.~\eqref{eq:three-point-open-current}.

\section{Higher-point derivation}
\label{app:higher-point}

Consider a connected tree diagram with $n\geq2$ photons. If the first
interaction adjacent to $u$ is cubic, the scalar line leaving it is
internal. Let this vertex lie at time $\eta$. The rest of the diagram
is a blob $\mathcal B(\eta')$ attached
through a scalar propagator, giving the profile
\begin{equation}
F(\eta)=\int_{-\infty}^{0}d\eta'\,
G_{P_{u\{a\}}}(\eta,\eta')\mathcal B(\eta').
\label{eq:endpoint-blob-profile}
\end{equation}
Here $P_{u\{a\}}=|\bp_u+\bk_a|$ is the merged scalar energy
defined in Eq.~\eqref{eq:endpoint-factors}.
All time integrals run from $-\infty$ to $0$, with the Bunch--Davies
prescription $\eta\to-\infty(1-i0)$ understood in the far past.
The blob includes all remaining vertices, external profiles, and
integrations. The cubic interaction is
\begin{equation}
\mathcal I_{\alpha\beta}[F]
:=i\int_{-\infty}^{0}d\eta\,e^{ik_a\eta}
F(\eta)V_{3,\alpha\beta}e^{ip_u\eta},
\label{eq:endpoint-vertex-integral}
\end{equation}
where $V_{3,\alpha\beta}$ is the differential operator recorded in
Appendix~\ref{app:vertices}, with derivatives acting on the adjacent scalar
profiles.

Integration by parts gives
\begin{equation}
\begin{aligned}
\mathcal I_{\alpha\beta}[F]
={}&-2\left(p_{u,\alpha\beta}-\frac{p_u}{k_a}k_{a,\alpha\beta}\right)
 i\int_{-\infty}^{0}d\eta\,e^{i(k_a+p_u)\eta}F(\eta)\\
&-\frac{k_{a,\alpha\beta}}{k_a}
 \left[e^{i(k_a+p_u)\eta}F(\eta)\right]_{-\infty}^{0}.
\end{aligned}
\label{eq:endpoint-boundary-identity}
\end{equation}
The derivative acting on $F$ has been transferred to the external
phase, leaving the displayed boundary term.
The Dirichlet condition $G_{P_{u\{a\}}}(0,\eta')=0$ implies $F(0)=0$
independently of the blob. Thus the contribution at $\eta=0$ vanishes;
the far-past contribution is suppressed by the prescribed contour.

The auxiliary spatial vectors defined in
Eq.~\eqref{eq:gauge-shift} satisfy
\begin{equation}
\begin{aligned}
\epsilon_a^{\alpha\beta}(u)
 \left(p_{u,\alpha\beta}-\frac{p_u}{k_a}k_{a,\alpha\beta}\right)&=0,\\
\epsilon_a^{\alpha\beta}(u)
(V_4)_{\alpha\beta,\gamma\delta}(a,b)
\epsilon_b^{\gamma\delta}(u)&=0.
\end{aligned}
\label{eq:common-reference-endpoint-identities}
\end{equation}
The first identity makes the integrated cubic interaction vanish.
For the quartic vertex in Appendix~\ref{app:vertices}, these vectors satisfy
$\epsilon_a^{\alpha\beta}\epsilon_{b,\alpha\beta}
=-(\epsilon_a^{\alpha\beta}k_{a,\alpha\beta})
(\epsilon_b^{\gamma\delta}k_{b,\gamma\delta})/(2k_a k_b)$,
where the argument $u$ is implicit. The two terms of $V_4$
therefore cancel. Every tree diagram begins at $u$ with either a cubic
or a quartic interaction, so each diagram vanishes separately. Hence
$\Psi^{\rm CR}_{\mathcal P_n}(u)=0$ for $n\geq2$.

At three points the other scalar is external, $F(0)=1$, and the boundary
term instead yields the nonzero seed
$\Psi^{\rm CR}_{\{a\}}(u)=k_a\df{a}{u}$ used in
Eq.~\eqref{eq:three-point-gauge-change}.

To recover the physical coefficient, expand each auxiliary vector as
$\epsilon_{a,\alpha\beta}(u)=\xi^{\hplus}_{a,\alpha\beta}
-\df{a}{u}k_{a,\alpha\beta}$
in the current contractions, giving
Eq.~\eqref{eq:longitudinal-subset-expansion}.
To evaluate the longitudinal contractions, we use the Abelian Ward
identity
\begin{equation}
\begin{aligned}
&k_a^{\alpha\beta}
\left\langle O_uJ_{a,\alpha\beta}
\prod_{b\in\mathcal P_n\setminus\{a\}}J_b^+O_v\right\rangle'\\
&\qquad=\Psi^\phi_{\mathcal P_n\setminus\{a\}}
 (\bp_{u},\bp_{v}+\bk_a)-\Psi^\phi_{\mathcal P_n\setminus\{a\}}
 (\bp_{u}+\bk_a,\bp_{v}).
\end{aligned}
\label{eq:abelian-ward}
\end{equation}
The two terms reflect the opposite scalar charges. Applying this identity
to every longitudinal insertion in
Eq.~\eqref{eq:longitudinal-subset-expansion} gives, for $n\geq2$,
\begin{equation}
\begin{aligned}
\Psi^\phi_{\mathcal P_n}
={}&\sum_{\varnothing\ne S\subseteq\mathcal P_n}
(-1)^{|S|+1}\prod_{a\in S}\df{a}{u}\sum_{L\sqcup R=S}(-1)^{|L|}
\Psi^\phi_{\mathcal P_n\setminus S}
(\bP_{u L},\bP_{v R}).
\end{aligned}
\label{eq:finite-gauge-relation}
\end{equation}
Here $S=L\sqcup R$, with $L$ and $R$ assigned to the $u$ and $v$
endpoints, respectively. Setting $C=\mathcal P_n\setminus S$ and using
$(-1)^{|S|+1+|L|}=(-1)^{|R|+1}$ gives
\begin{equation}
\begin{aligned}
\Psi^\phi_{\mathcal P_n}
={}&\sum_{\substack{L\sqcup C\sqcup R=\mathcal P_n\\
L\cup R\ne\varnothing}}
(-1)^{|R|+1}
\left(\prod_{a\in L\cup R}\df{a}{u}\right)
\Psi^\phi_C(\bP_{uL},\bP_{vR}).
\end{aligned}
\label{eq:three-partition-ward}
\end{equation}

Every coefficient on the right has fewer photons. Thus the two-point
normalization and the three-point result
Eq.~\eqref{eq:three-point-gauge-change} determine all higher coefficients
recursively, with the scalar energies and spinors evaluated at the
shifted momenta at each step. The explicit factors in
Eq.~\eqref{eq:three-partition-ward}, however, retain the original reference
spinor $u$.

\subsection{Endpoint symmetrization}

We now convert the single-reference Ward recursion into
Eq.~\eqref{eq:scalar-master}. Define
\begin{equation}
 \delta_a:=\df{a}{u}-\df{a}{v}
 =\frac{\ang{u}{v}}{\ang{u}{a}\ang{v}{a}}.
 \label{eq:delta-endpoint}
\end{equation}
We work at generic momenta where the displayed denominators are nonzero.

Before imposing the common-reference zero, the subset expansion gives
\begin{equation}
\Psi^{\rm CR}_{K}(u;\bp_u,\bp_v)
=\sum_{L\sqcup C\sqcup R=K}(-1)^{|R|}
\prod_{a\in L\cup R}\df{a}{u}\,
\Psi^\phi_C(\bP_{uL},\bP_{vR}).
\label{eq:cr-three-partition}
\end{equation}
Here the scalar momenta obey
$\bp_u+\bp_v+\bk_K=0$, and the reference spinor belongs to the
actual endpoint $\bp_u$. For $|K|\geq2$, the left-hand side vanishes;
moving the term $L=R=\varnothing$ to the other side then recovers
Eq.~\eqref{eq:three-partition-ward}.

\paragraph{Localization of the partition difference.}
For $n\geq2$, let
\begin{equation}
\Delta_n:=\Psi^\phi_{\mathcal P_n}
-\!\!\sum_{\substack{L\sqcup C\sqcup R=\mathcal P_n\\L\cup R\ne\varnothing}}
(-1)^{|R|+1}
\prod_{a\in L}\df{a}{u}\prod_{b\in R}\df{b}{v}\,
\Psi^\phi_C(\bP_{uL},\bP_{vR}).
\label{eq:partition-difference-target}
\end{equation}
Using Eq.~\eqref{eq:three-partition-ward} gives
\begin{equation}
\Delta_n=
\sum_{\substack{L\sqcup C\sqcup R=\mathcal P_n\\R\ne\varnothing}}
(-1)^{|R|+1}\prod_{a\in L}\df{a}{u}
\left[\prod_{b\in R}\df{b}{u}-\prod_{b\in R}\df{b}{v}\right]
\Psi^\phi_C(\bP_{uL},\bP_{vR}).
\label{eq:partition-difference-first}
\end{equation}
Expand the bracket using $D_{a|v}=D_{a|u}-\delta_a$:
\begin{equation}
\prod_{b\in R}\df{b}{u}-\prod_{b\in R}\df{b}{v}
=\sum_{\varnothing\ne T\subseteq R}
(-1)^{|T|+1}
\prod_{a\in T}\delta_a\prod_{b\in R\setminus T}\df{b}{u}.
\label{eq:product-difference}
\end{equation}
Writing $R=T\sqcup R'$ then yields
\begin{equation}
\Delta_n=
\sum_{\varnothing\ne T\subseteq\mathcal P_n}
\left(\prod_{a\in T}\delta_a\right)
\sum_{L\sqcup C\sqcup R'=\mathcal P_n\setminus T}
(-1)^{|R'|}\prod_{b\in L\cup R'}\df{b}{u}\,
\Psi^\phi_C\!\left(\bP_{uL},\bP_{v(T\cup R')}\right).
\label{eq:reindexed-partition-difference}
\end{equation}

The inner sum is Eq.~\eqref{eq:cr-three-partition} for the remaining
photons, with scalar momenta $(\bp_u,\bP_{vT})$. Only the $v$ endpoint
has been shifted: the reference spinor $u$ still belongs to the unchanged
endpoint, so the common-reference zero applies. Thus
\begin{equation}
\Delta_n=
\sum_{T\subseteq\mathcal P_n}
\left(\prod_{a\in T}\delta_a\right)
\Psi^{\rm CR}_{\mathcal P_n\setminus T}
\!\left(u;\bp_u,\bP_{vT}\right),
\label{eq:cr-localization}
\end{equation}
where the added $T=\varnothing$ term vanishes for $n\geq2$.
All terms with $|\mathcal P_n\setminus T|\geq2$ also vanish. The case
$T=\mathcal P_n$ gives $p_u\prod_a\delta_a$, since
$\bP_{v\mathcal P_n}=-\bp_u$. The remaining cases,
$T=\mathcal P_n\setminus\{a\}$, give the one-photon auxiliary seed
$k_aD_{a|u}$. Therefore
\begin{equation}
\Delta_n=
\left(\prod_{a\in\mathcal P_n}\delta_a\right)
\left[p_u+\sum_{a\in\mathcal P_n}
 k_a\frac{\df{a}{u}}{\delta_a}\right].
\label{eq:partition-identity-I}
\end{equation}

\paragraph{Evaluation of the kinematic factor.}
Equation~\eqref{eq:delta-endpoint} implies
\begin{equation}
 k_a\frac{\df{a}{u}}{\delta_a}
 =\frac{\ang{u}{\bar a}\ang{v}{a}}{2\ang{u}{v}}.
 \label{eq:kinematic-ratio}
\end{equation}
Schouten's identity and $\ang{a}{\bar a}=-2k_a$ give
\begin{equation}
 \ang{u}{\bar a}\ang{v}{a}-\ang{u}{a}\ang{v}{\bar a}
 =2k_a\ang{u}{v}.
 \label{eq:kinematic-schouten}
\end{equation}
Contracting spatial momentum conservation with
$\lambda_u^\alpha\lambda_v^\beta$ gives the complementary sum,
\begin{equation}
 \sum_{a\in\mathcal P_n}
 \left[\ang{u}{\bar a}\ang{v}{a}+\ang{u}{a}\ang{v}{\bar a}\right]
 =2(p_v-p_u)\ang{u}{v}.
 \label{eq:kinematic-momentum-conservation}
\end{equation}
Adding the summed Schouten identity to this equation yields
\begin{equation}
 \sum_{a\in\mathcal P_n}\ang{u}{\bar a}\ang{v}{a}
 =\left(p_v-p_u+\sum_{a\in\mathcal P_n}k_a\right)\ang{u}{v}.
 \label{eq:kinematic-sum}
\end{equation}
Together with Eq.~\eqref{eq:kinematic-ratio}, this proves
\begin{equation}
 p_u+\sum_{a\in\mathcal P_n}
 k_a\frac{\df{a}{u}}{\delta_a}
 =\frac{\ET}{2}.
\label{eq:partition-identity-II}
\end{equation}

Substituting into Eq.~\eqref{eq:partition-identity-I}, we obtain
\begin{equation}
\Delta_n=\frac{\ET}{2}\prod_{a\in\mathcal P_n}
\left(\df{a}{u}-\df{a}{v}\right)
=\frac{\ET\ang{u}{v}^{n}}
{2\prod_{a=1}^{n}\ang{u}{a}\ang{v}{a}}.
\label{eq:partition-identity-final}
\end{equation}
This is the term required to relate
Eq.~\eqref{eq:partition-difference-target} to
Eq.~\eqref{eq:scalar-master}. The reorganization above applies for
$n\geq2$; the three-point seed in
Eq.~\eqref{eq:three-point-gauge-change} supplies $n=1$.


\subsection{Four-point specialization}
\label{app:four-point-specialization}

For two photons, the eight nontrivial partitions in
Eq.~\eqref{eq:scalar-master} give
\begin{equation}
\begin{aligned}
 \Psi^\phi_{\{1,2\}}
 ={}&\frac{\ET\ang{u}{v}^{2}}
 {2\ang{u}{1}\ang{v}{1}\ang{u}{2}\ang{v}{2}}
 -\df{1}{u}\df{2}{u}p_v-\df{1}{v}\df{2}{v}p_u\\
 &+\Bigl[
 -\df{1}{u}\Psi^\phi_{\{2\}}(\bP_{u1},\bp_v)
 +\df{1}{v}\Psi^\phi_{\{2\}}(\bp_u,\bP_{v1})\\
 &\qquad+\df{1}{u}\df{2}{v}P_{u1}
 +(1\leftrightarrow2)\Bigr].
\end{aligned}
\label{eq:four-point-partition-expansion}
\end{equation}
The two-point terms use $P_{u12}=p_v$ and $P_{v12}=p_u$.
The three-point terms can be evaluated in the forms
\begin{equation}
\begin{aligned}
 \Psi^\phi_{\{2\}}(\bP_{u1},\bp_v)
 &=D_{2|u1}(P_{u1}+k_2-p_v),\\
 \Psi^\phi_{\{2\}}(\bp_u,\bP_{v1})
 &=-D_{2|v1}(P_{v1}+k_2-p_u).
\end{aligned}
\end{equation}
Here the labels $u1,v1$ on the $D$ factors denote merged endpoint
spinors. Substituting these expressions into
Eq.~\eqref{eq:four-point-partition-expansion}, using
$P_{v1}=P_{u2}$, and applying momentum conservation and Schouten's
identity yields Eq.~\eqref{eq:scalar-four-point}.


\section{Conformal Ward identity}
\label{app:symmetry}

We verify Eq.~\eqref{eq:scalar-CWI} for the scalar formula
\eqref{eq:scalar-master}. The special conformal generator is
\begin{equation}
 \mathcal K^i
 =2(\sigma^i)_\alpha{}^\beta
 \sum_j\frac{\partial^2}
 {\partial\lambda_{j\alpha}\partial\bar\lambda_j{}^\beta},
 \label{eq:conformal-generator}
\end{equation}
where the sum runs over both scalar endpoints and all photons.

\paragraph{Endpoint intertwiner.}
Let $r\in\{u,v\}$ and $D_{S|r}:=\prod_{a\in S}D_{a|r}$.
For a function $F(\bP_{rS})$, we use the endpoint identity
\begin{equation}
\begin{aligned}
 \mathcal K^i_{\{S,r\}}
 [D_{S|r}F(\bP_{rS})]
 ={}&D_{S|r}\mathcal K^i_{P_{rS}}F(\bP_{rS})\\
 &+2\sum_{a\in S}\frac{\xi_a^{\hplus,i}}{k_a^2}
 D_{S\setminus\{a\}|r}F(\bP_{rS}).
\end{aligned}
\label{eq:D-intertwiner-main}
\end{equation}
Here $\mathcal K^i_{\{S,r\}}$ acts on the endpoint $r$ and the photons
in $S$. We use this identity below to organize the action of
$\mathcal K^i$ on the partition sum.

\paragraph{The explicit total-energy term.}
Write the first term of Eq.~\eqref{eq:scalar-master} as
$E_{\rm T}H_n$, with
\[
 H_n:=\frac{\ang{u}{v}^{n}}
 {2\prod_{a\in\mathcal P_n}\ang{u}{a}\ang{v}{a}},
 \qquad E_{\rm T}=-\frac12\sum_j\ang{j}{\bar j}.
\]
Since $H_n$ is holomorphic, a barred-spinor derivative acts only on
$E_{\rm T}$. If the remaining derivative also acts on $E_{\rm T}$,
the result vanishes by $\operatorname{tr}\sigma^i=0$. Otherwise the
result is proportional to
\[
 (\sigma^i)_\alpha{}^\beta
 \sum_j\lambda_{j\beta}
 \frac{\partial H_n}{\partial\lambda_{j\alpha}}=0.
\]
The last equality follows because all angle brackets, and hence $H_n$,
are invariant under a simultaneous $SL(2,\mathbb C)$ transformation of
the unbarred spinors; its infinitesimal generators are traceless.
Thus $\mathcal K^i(E_{\rm T}H_n)=0$. The partition sum must supply the
inhomogeneous CWI source.

\paragraph{Induction and layer cancellation.}
The induction uses the homogeneous CWI of the scalar two-point
coefficient. With one photon, the formula reads
\[
 \Psi^\phi_{\{a\}}(\bp_u,\bp_v)
 =E_{\rm T}H_1
 -D_{a|u}\Psi^\phi_\varnothing(\bP_{u\{a\}},\bp_v)
 +D_{a|v}\Psi^\phi_\varnothing(\bp_u,\bP_{v\{a\}}).
\]
Using the endpoint intertwiner and the homogeneous two-point CWI,
together with $\mathcal K^i(E_{\rm T}H_1)=0$, gives
Eq.~\eqref{eq:scalar-CWI} for this base case.

Assume the CWI holds for fewer than $n$ photons. Every partition in
Eq.~\eqref{eq:scalar-master} has $|C|<n$, so
\begin{equation}
\begin{aligned}
 &\left(\mathcal K^i_{P_{uL}}+\mathcal K^i_{P_{vR}}
       +\sum_{c\in C}\mathcal K_c^i\right)
 \Psi_C^\phi(\bP_{uL},\bP_{vR})\\
 &\quad=2\sum_{a\in C}\frac{\xi_a^{\hplus,i}}{k_a^2}
 \Big[
 \Psi^\phi_{C\setminus\{a\}}
 (\bP_{uL},\bP_{v,R\cup\{a\}})\\
 &\hspace{8em}-\Psi^\phi_{C\setminus\{a\}}
 (\bP_{u,L\cup\{a\}},\bP_{vR})\Big].
\end{aligned}
\label{eq:induction-CWI}
\end{equation}
For compactness, define
\[
 W_{L,R}:=(-1)^{|R|+1}D_{L|u}D_{R|v},
 \qquad
 \sum_{\Pi_s}:=\sum_{\substack{L\sqcup C\sqcup R=\mathcal P_n\\
                       |L|+|R|=s}}.
\]
Applying the intertwiner at both endpoints gives
\begin{equation}
 \mathcal K^i\!\left[
 \sum_{\Pi_s} W_{L,R}\Psi_C^\phi(\bP_{uL},\bP_{vR})\right]
 =\mathcal I_s^i+\mathcal E_s^i,
 \label{eq:layer-CWI-decomposition}
\end{equation}
where the lower-point CWI contributes
\begin{equation}
\begin{aligned}
 \mathcal I_s^i
 :={}&2\sum_{\Pi_s} W_{L,R}
 \sum_{a\in C}\frac{\xi_a^{\hplus,i}}{k_a^2}
 \Big[\Psi^\phi_{C\setminus\{a\}}
 (\bP_{uL},\bP_{v,R\cup\{a\}})\\
 &\hspace{5em}-\Psi^\phi_{C\setminus\{a\}}
 (\bP_{u,L\cup\{a\}},\bP_{vR})\Big],
\end{aligned}
\label{eq:induced-layer-source}
\end{equation}
and the differentiated endpoint factors contribute
\begin{equation}
\begin{aligned}
 \mathcal E_s^i
 :={}&2\sum_{\Pi_s}(-1)^{|R|+1}\Bigg[
 \sum_{a\in L}\frac{\xi_a^{\hplus,i}}{k_a^2}
 D_{L\setminus\{a\}|u}D_{R|v}\\
 &\hspace{4em}+\sum_{a\in R}\frac{\xi_a^{\hplus,i}}{k_a^2}
 D_{L|u}D_{R\setminus\{a\}|v}\Bigg]
 \Psi_C^\phi(\bP_{uL},\bP_{vR}).
\end{aligned}
\label{eq:endpoint-layer-source}
\end{equation}

To compare neighboring layers, mark the differentiated photon $a$ in a
layer-$(s+1)$ partition $(L',C',R')$. If $a\in L'$, relabel
$(L,C,R)=(L'\setminus\{a\},C'\cup\{a\},R')$; if $a\in R'$, use
$(L,C,R)=(L',C'\cup\{a\},R'\setminus\{a\})$. Each relabeling is a
bijection onto layer-$s$ partitions with a marked $a\in C$. The two
contributions are
\begin{equation*}
\begin{aligned}
 \mathcal E_{s+1,L}^i
 ={}&2\sum_{\Pi_s} W_{L,R}\sum_{a\in C}
 \frac{\xi_a^{\hplus,i}}{k_a^2}
 \Psi^\phi_{C\setminus\{a\}}
 (\bP_{u,L\cup\{a\}},\bP_{vR}),\\
 \mathcal E_{s+1,R}^i
 ={}&-2\sum_{\Pi_s} W_{L,R}\sum_{a\in C}
 \frac{\xi_a^{\hplus,i}}{k_a^2}
 \Psi^\phi_{C\setminus\{a\}}
 (\bP_{uL},\bP_{v,R\cup\{a\}}).
\end{aligned}
\end{equation*}
The left-endpoint contribution cancels the second term in
$\mathcal I_s^i$. The right-endpoint contribution cancels the first:
its sign changes because $|R'|=|R|+1$. Hence
\begin{equation}
 \mathcal E_{s+1}^i=-\mathcal I_s^i,
 \qquad s=1,\ldots,n-1.
 \label{eq:neighboring-layer-cancellation}
\end{equation}
Since $C=\varnothing$ in the last layer, the sum defining
$\mathcal I_n^i$ is empty. The full partition sum therefore telescopes:
\begin{equation}
\begin{aligned}
 &\mathcal K^i\!\left[
 \sum_{s=1}^n\sum_{\Pi_s} W_{L,R}
 \Psi_C^\phi(\bP_{uL},\bP_{vR})\right]\\
 &\quad=\mathcal E_1^i
 +\sum_{s=1}^{n-1}(\mathcal I_s^i+\mathcal E_{s+1}^i)
 =\mathcal E_1^i.
\end{aligned}
\label{eq:layer-telescoping}
\end{equation}
Only one photon is absorbed in the first layer, giving
\begin{equation}
\begin{aligned}
 \mathcal E_1^i
 ={}&2\sum_{a\in\mathcal P_n}\frac{\xi_a^{\hplus,i}}{k_a^2}
 \Big[\Psi^\phi_{\mathcal P_n\setminus\{a\}}
 (\bp_u,\bP_{v\{a\}})\\
 &\hspace{6em}-\Psi^\phi_{\mathcal P_n\setminus\{a\}}
 (\bP_{u\{a\}},\bp_v)\Big].
\end{aligned}
\label{eq:first-layer-source}
\end{equation}
Together with $\mathcal K^i(E_{\rm T}H_n)=0$, the stated endpoint
intertwiner and homogeneous two-point CWI yield
Eq.~\eqref{eq:scalar-CWI} for Eq.~\eqref{eq:scalar-master}.
This is a consistency check, not a uniqueness argument: the homogeneous
term is not fixed by this Ward identity alone.

\end{document}